\documentclass[preprintnumbers,amsmath,amssymb,showpacs]{revtex4}

\usepackage{graphicx}
\usepackage{dcolumn}
\usepackage{bm}
\usepackage{subfigure}
\usepackage[                   
            pdfstartview=FitH,
            colorlinks, 
            pdfborder=001,   
            linkcolor=blue,
            anchorcolor=blue,
            citecolor=blue,
            urlcolor=blue
            ]{hyperref}

\begin{document}

\title{Detection of multipartite entanglement via quantum Fisher information }

\author{Yan Hong}
 \affiliation {School of Mathematics and Science, Hebei GEO University, Shijiazhuang 050031,  China}

\author{Xianfei Qi}
 \affiliation {School of Mathematics and Statistics, Shangqiu Normal University, Shangqiu 476000,  China}

\author{Ting Gao}
\email{gaoting@hebtu.edu.cn} \affiliation {School of Mathematical Sciences, Hebei Normal University, Shijiazhuang 050024,  China}

\author{Fengli Yan}
\email{flyan@hebtu.edu.cn} \affiliation {College of Physics, Hebei Normal University, Shijiazhuang 050024,  China}

\begin{abstract}
In this paper, we  focus on two different kinds of multipartite  correlation, $k$-nonseparability and $k$-partite entanglement, both of which can describe the essential characteristics of multipartite entanglement.
We propose effective methods to detect $k$-nonseparability and $k$-partite entanglement in terms of quantum Fisher information. We illustrate the significance of our results  and show that they identify some $k$-nonseparability and $k$-partite entanglement that cannot be identified by known criteria by several concrete examples.

\end{abstract}


\maketitle

\section{Introduction}
Quantum entanglement is one of the most fundamental features in composite quantum systems that has no classical counterpart. In recent years, it was identified as the key resource in a variety of applications ranging from quantum dense coding \cite{PRL69.2881}, quantum teleportation \cite{PRL70.1895,EPL84.50001}, quantum computation \cite{Nature2000}, to quantum metrology \cite{Nature2010}.

An fundamental problem in quantum entanglement theory is to decide whether a given quantum state is entangled or separable. For bipartite systems, much progress has been made on the criteria for the separability of mixed states \cite{RMP81.865,PR474.1}. However, the situation becomes more complicated when the number of parties increases. For example, there are only two cases in bipartite systems, separable or entangled, while, in multipartite quantum systems, there exist several inequivalent classes of multiparticle entanglement and it is difficult to decide to which class a given state belongs. A classification of entanglement for general multipartite systems was proposed according to the question:\textquotedblleft How many partitions are separable?\textquotedblright and therefore quantum states can be classified into $k$-separable states and $k$-nonseparable states. The detection of $k$-nonseparability has been investigated extensively, many efficient criteria \cite{QIC2008,QIC2010,PRA82.062113,EPL104.20007,PRA91.042313,SR5.13138,PRA93.042310,SciChina2017} and computable measures \cite{PRA68.042307,PRL93.230501,PRA83.062325,PRA86.062323,PRL112.180501} have been presented. Alternatively, another classification was proposed according to the question:\textquotedblleft How many partitions are entangled?\textquotedblright and the concepts of $k$-producible state and $(k + 1)$-partite entangled state are naturally defined. $k$-partite entangled state and $k$-nonseparable state are two different concepts involving the partitions of subsystem in $N$-partite quantum systems, although $2$-nonseparable state is the same as $N$-partite entangled state (i.e. genuine entangled state).

Quantum Fisher information (QFI) is the cornerstone of modern quantum metrology and plays an important role in determining the reachable accuracy of the estimated quantity \cite{PRL72.3439}. The QFI has been also used in the detection of the entanglement \cite{Hyllus2012,Toth2012,PRA88.014301,PRA99.012304} and the $k$-nonseparability of multipartite quantum systems \cite{PRA91.042313}. In this paper, we investigate the detection of entanglement for arbitrary multipartite quantum systems. We will propose effective methods to detect $k$-nonseparability and $k$-partite entanglement by making use of QFI. Several typical examples are provided to show that our results are stronger than previously reported ones.

The remainder of the paper is arranged as follows. In Sec.~\uppercase\expandafter{\romannumeral 2}, some preliminaries about $k$-separable state, $k$-producible state and quantum Fisher information are reviewed. In Sec.~\uppercase\expandafter{\romannumeral 3}, we derive effective criteria based on QFI, and provide examples to illustrate the efficiency of our results. Finally, a brief summary is given in Sec.~\uppercase\expandafter{\romannumeral 4}.

\section{Preliminaries}
We first review some elementary concepts that will be used in the rest of the paper. For an $N$-partite quantum system with Hilbert space $\mathcal{H}=\mathcal{H}_1\otimes \mathcal{H}_2\otimes\cdots \otimes\mathcal{H}_N$, where the dimension of $\mathcal{H}_i$ is denoted by $d_i$, an $N$-partite pure state $|\psi\rangle\in \mathcal{H}$ is called $k$-separable if there is a partition $\{\alpha_1,\ldots,\alpha_k\}$ with $\alpha_j$ being disjoint and $\bigcup\limits_{j=1}^k\alpha_j=\{1,2,\cdots,N\}$, such that it can be written as
$$|\psi\rangle=\bigotimes\limits_{j=1}^k|\psi_{\alpha_j}\rangle.$$
An $N$-partite mixed state is called $k$-separable,
 if it can be written as a convex combination of $k$-separable pure states
 $$\rho=\sum\limits_mp_m|\psi^{(m)}\rangle\langle\psi^{(m)}|,$$
 where pure state $|\psi^{(m)}\rangle$ might be $k$-separable under different partitions.
 If quantum state $\rho$ is not $k$-separable, then  it  is $k$-nonseparable. Here $2 \leq k\leq N$.

It is useful for the following to introduce the notion of $k$-producible states. For an $N$-partite quantum system with Hilbert space $\mathcal{H}=\mathcal{H}_1\otimes \mathcal{H}_2\otimes\cdots \otimes\mathcal{H}_N$, we call pure state $|\psi\rangle$ $k$-producible if
$|\psi\rangle=\bigotimes\limits_{j=1}^t|\psi_{\alpha_j}\rangle$  with $|\psi_{\alpha_j}\rangle$ being partial state with particles less than $k$.
A mixed state $\rho$ is called $k$-producible if $\rho=\sum\limits_mp_m|\psi^{(m)}\rangle\langle\psi^{(m)}|$ with pure state $|\psi^{(m)}\rangle$ being $k$-producible. If quantum state $\rho$ is not $k$-producible,  then we say it contains $(k+1)$-partite entanglement. Here $1 \leq k\leq N-1$.

We now introduce the notion related to QFI. For a quantum state $\rho$ and an observable $A$, the variance $V(\rho,A)$ and quantum Fisher information $F(\rho,A)$ are defined as following,
$$\begin{array}{ll}
V(\rho,A)=&\textrm{tr}\rho A^2-(\textrm{tr}\rho A)^2,\\
F(\rho,A)=&\sum\limits_{l,l'}\dfrac{(\lambda_l-\lambda_{l'})^2}{2(\lambda_l+\lambda_{l'})}|\langle l|A|l'\rangle|^2,
\end{array}$$
with $\rho$ having the spectral decomposition $\rho=\sum\limits_l\lambda_l|l\rangle\langle l|$.

It's worth noting that QFI reduces to variance for pure states, that is, $V(\rho,A)=F(\rho,A)$ for any pure state $\rho$.  In addition, quantum Fisher information possesses some important properties which play a key role in the following.

(1) Convexity:
$$F\left(\sum\limits_mp_m\rho_m,A\right)\leq\sum\limits_mp_mF(\rho_m,A),$$
where $\sum\limits_mp_m=1$.

(2) Additivity:
$$F\left(\bigotimes\limits_{i=1}^N\rho_i,\sum\limits_{i=1}^NH_i\right)=\sum\limits_{i=1}^NF(\rho_i,h_i),$$
where $H_i=\textbf{I}_1\otimes\cdots \otimes \textbf{I}_{i-1}\otimes h_i\otimes \textbf{I}_{i+1}\otimes\cdots \otimes\textbf{I}_N$
with $h_i$ being an observable  on $i$-th subsystem and $\textbf{I}_j$ being identity matrix on $j$-th subsystem.

\section{ multipartite entanglement criteria via quantum Fisher information}

For the convenience of the following statements, we first introduce some symbols.
In an $N$-partite quantum system with Hilbert space $\mathcal{H}=\mathcal{H}_1\otimes \mathcal{H}_2\otimes\cdots \otimes\mathcal{H}_N$, for any  nonempty subset $\alpha$ of $\{1,2,\cdots,N\}$,
let $\rho_\alpha$ be partial state of $\rho$ on subsystem $\bigotimes\limits_{i\in\alpha}\mathcal{H}_i$, $A_i$ be any Hermite operator acting on $\mathcal{H}_i$, and
$A_\alpha=\sum\limits_{i\in\alpha}A_i$  acting on $\bigotimes\limits_{i\in\alpha}\mathcal{H}_i$.

Now based on the above concepts, we establish the following criterion.

 \emph{Theorem.} For an $N$-partite quantum system with Hilbert space $\mathcal{H}=\mathcal{H}_1\otimes \mathcal{H}_2\otimes\cdots \otimes\mathcal{H}_N$,
 if
 \begin{equation}\label{sc1}
\begin{array}{rl}
F(\rho_\alpha,A_\alpha)\leq F_\alpha
\end{array}
\end{equation}
 where $F_\alpha$ is the constant only depending on subset $\alpha$, then \\

(I) for any $k$-separable quantum state $\rho$, we have
\begin{equation}\label{kseparable}
\begin{array}{rl}
F(\rho,A)\leq \textrm{min} \sum\limits_{l=1}^kF_{\alpha_l},
\end{array}
\end{equation}
where $A=\sum\limits_{i=1}^NA_i$ and the minimum min takes over all possible partitions $\{\alpha_1,\alpha_2,\cdots,\alpha_k\}$ of $\{1,2,\cdots,N\}$.\\

(II) for any $k$-producible quantum state $\rho$, one has
\begin{equation}\label{kproducible}
\begin{array}{rl}
F(\rho,A)\leq \textrm{min}' \sum\limits_{l=1}^tF_{\alpha_l},
\end{array}
\end{equation}
where $A=\sum\limits_{i=1}^NA_i$ and the minimum min$'$ takes over all possible partitions $\{\alpha_1,\alpha_2,\cdots,\alpha_t\}$ of $\{1,2,\cdots,N\}$ with the number of particles in $\alpha_l$  being less than $k$.

\emph{Proof.} Let us prove (I) holds for all $k$-separable states. Suppose that $\rho=|\psi\rangle\langle\psi|$ is $k$-separable pure state under partition $\{\alpha_1,\cdots,\alpha_k\}$, that is,
$|\psi\rangle=\bigotimes\limits_{l=1}^k|\psi_{\alpha_l}\rangle$, then
$$\begin{array}{ll}
F(|\psi\rangle,A)&=\sum\limits_{l=1}^kF(|\psi_{\alpha_l}\rangle,A_{\alpha_l})\\
&\leq\sum\limits_{l=1}^kF_{\alpha_l},
\end{array}$$
where equality follows from additivity of quantum Fisher information, and inequality holds by inequality  (\ref{sc1}).
Hence, we have
\begin{equation}\label{ksp}
\begin{array}{rl}
F(|\psi\rangle,A)\leq \textrm{min}\sum\limits_{l=1}^kF_{\alpha_l},
\end{array}
\end{equation}
the minimum takes over all possible partitions $\{\alpha_1,\alpha_2,\cdots,\alpha_k\}$ of $\{1,2,\cdots,N\}$.

Let  $\rho=\sum\limits_mp_m|\psi^{(m)}\rangle\langle\psi^{(m)}|$ be $k$-separable mixed state where $\sum\limits_mp_m=1$ and $|\psi^{(m)}\rangle$ is
$k$-separable pure state, then
$$\begin{array}{ll}
F(\rho,A)&\leq\sum\limits_mp_mF(|\psi^{(m)}\rangle,A)\\
&\leq\sum\limits_mp_m\textrm{min}\sum\limits_{l=1}^kF_{\alpha_l}\\
&=\textrm{min}\sum\limits_{l=1}^kF_{\alpha_l},
\end{array}$$
where the minimum takes over all possible partitions $\{\alpha_1,\alpha_2,\cdots,\alpha_k\}$ of $\{1,2,\cdots,N\}$.
Here, the first inequality is true because of convexity of quantum Fisher information, the second inequality holds by inequality (\ref{ksp}), and the equality is valid from $\sum\limits_mp_m=1$. So far we have proved the validity of the inequality (\ref{kseparable}) for any $k$-separable quantum state. Similarly, we can prove (II) holds for any $k$-producible quantum state.

In order to get more practical results,  let's consider the special case of the Theorem.
For an $N$-partite quantum system with Hilbert space $\mathcal{H}=\mathcal{H}_1\otimes \mathcal{H}_2\otimes\cdots \otimes\mathcal{H}_N$ where the dimension of  $\textrm{dim} \mathcal{H}_i$ is  $d$ for $1\leq i\leq N$,
let
\begin{equation}\label{Ai}
\begin{array}{rl}
A_i=\sum\limits_{j=1}^na_{ij}A_i^{(j)}
\end{array}
\end{equation}
  where $a_{ij}$ is a real number  satisfying  $\sum\limits_{j=1}^na_{ij}^2=1$, and $\{A_i^{(1)},\cdots,A_i^{(n)}\}$ is the set consisting of
 arbitrary $n$ anticommuting Hermite operators acting on $\mathcal{H}_i$ with $(A_i^{(j)})^2\leq \textbf{I}$  for all $i$.
For these particular operators, one has $A_\alpha^2\leq n_\alpha^2\bigotimes\limits_{i\in\alpha}\textbf{I}_i$  with $n_\alpha$ being the number of particles of the set $\alpha$ \cite{PRA91.042313}.  And then we get
$$\begin{array}{ll}
F(\rho_\alpha,A_\alpha)&\leq\sum\limits_mp_mF(|\psi^{(m)}_\alpha\rangle,A_\alpha)\\
&=\sum\limits_mp_mV(|\psi^{(m)}_\alpha\rangle,A_\alpha)\\
&=\sum\limits_mp_m\left[\langle\psi^{(m)}_\alpha| A_\alpha^2|\psi^{(m)}_\alpha\rangle-\Big(\langle\psi^{(m)}_\alpha|A_\alpha|\psi^{(m)}_\alpha\rangle\Big)^2\right]\\
&\leq\sum\limits_mp_m\langle\psi^{(m)}_\alpha| A_\alpha^2|\psi^{(m)}_\alpha\rangle\\
&\leq n_\alpha^2,
\end{array}$$
where $\rho_\alpha=\sum\limits_mp_m|\psi^{(m)}_\alpha\rangle\langle\psi^{(m)}_\alpha|$.

Based on the above special case, we further obtain the following conclusion:

\emph{Corollary.} For an $N$-partite quantum system with Hilbert space $\mathcal{H}=\mathcal{H}_1\otimes \mathcal{H}_2\otimes\cdots \otimes\mathcal{H}_N$ and Dim$\mathcal{H}_i=d$ for $1\leq i\leq N$, let $A=\sum\limits_{i=1}^NA_i$ with  $A_i$ being defined by (\ref{Ai}),\\

(I) if $\rho$ is $k$-separable, then
$$F(\rho,A)\leq\left\{
\begin{array}{ll}
   v(u+1)^2+(k-v)u^2, &\textrm{ for }k\nmid N \\
   ku^2,&\textrm{ for } k|N\\
\end{array}\right.$$
where   $u=[\frac{N}{k}],v=N-ku$.\\

(II) if $\rho$ is $k$-producible, then
\begin{equation}\label{}
\begin{array}{rl}
F(\rho,A)\leq k^2+N-k.
\end{array}
\end{equation}

$\emph{Example 1}.$ Consider the $8$-qubit mixed states,
\begin{eqnarray*}
\rho(p,q)=(1-p)|G\rangle\langle G|+q|\widetilde{G}\rangle\langle \widetilde{G}|+\frac{p-q}{2^{8}}\textbf{1},
\end{eqnarray*}
where $|G\rangle=\frac{1}{\sqrt{2}}(|0\rangle^{\otimes8}+|1\rangle^{\otimes8}),
|\widetilde{G}\rangle=\frac{1}{\sqrt{2}}(|0\rangle^{\otimes8}-\textrm{i}|1\rangle^{\otimes8}).$

In our corollary, let $\{A_i^{(1)},\cdots,A_i^{(n)}\}=\{\sigma_x,\sigma_y,\sigma_z\} $ and $A_i=\sigma_z$ for $1\leq i\leq 8$.
For these  mixed states $\rho(p,q)$, the parameter ranges of 3-nonseparability and 5-nonseparability identified by the theorem of Ref.\cite{QIC2010} are illustrated in Fig.1.
We clearly find that these  3-nonseparable states in the
area enclosed by red line $a$, line $p=1$, blue line $b$ and $p$ axis only are
detected by  our corollary, but not by the theorem of Ref.\cite{QIC2010}. Similarly, these 6-nonseparable states in the
area enclosed by   purple line $c$, line $p=1$, green line $d$ and $p$ axis only are
detected by  our corollary, but not by the theorem of Ref.\cite{QIC2010}.

\begin{figure}
\begin{center}
  \subfigure{
    \label{}
    \includegraphics[scale=0.6]{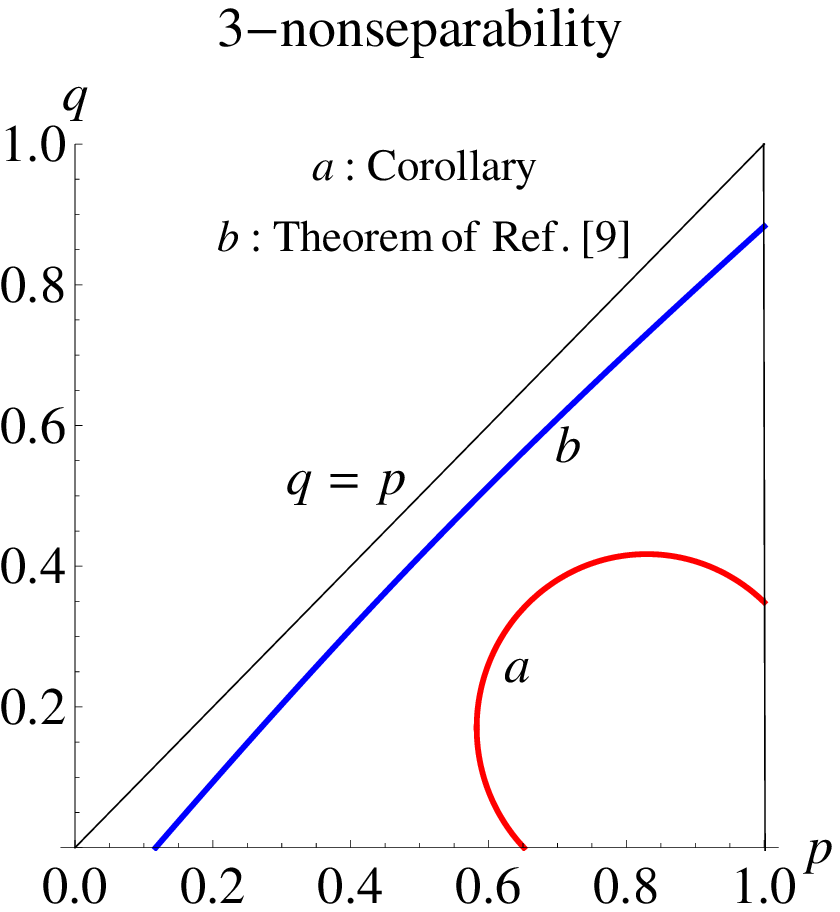}}
  \hspace{0.2in}
  \subfigure{
    \label{}
    \includegraphics[scale=0.6]{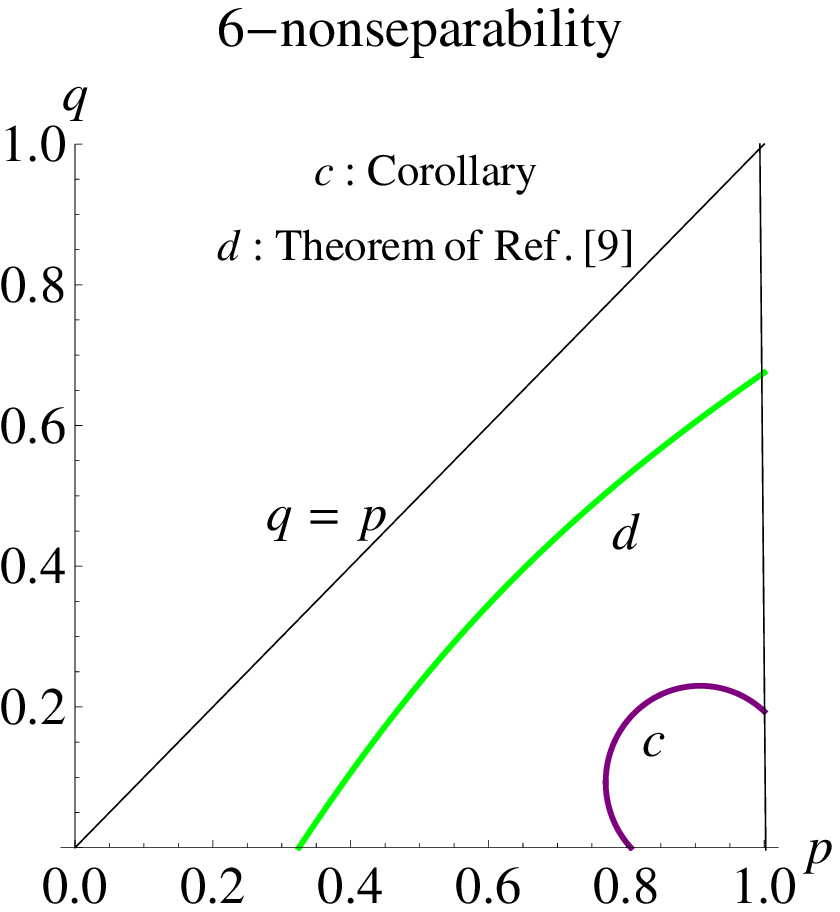}}\\
\caption[\label{}
]{ Comparison of  detection ability between
our corollary  and theorem of Ref.\cite{QIC2010} for the quantum states
$\rho(p,q)=(1-p)|G\rangle\langle G|+q|\widetilde{G}\rangle\langle \widetilde{G}|+\frac{p-q}{2^{8}}\textbf{1}$
when $k=3$  and $k=5$. These quantum states in area enclosed by
red line $a$, $p=1$,  line $q=p$ and $p$ axis are
3-nonseparable states detected by our corollary; these quantum states in area enclosed by
blue line $b$, line $p=1$,  line $q=p$ and $p$ axis are
3-nonseparable states detected by the theorem of Ref.\cite{QIC2010}. These  3-nonseparable states in the
area enclosed by red line $a$, line $p=1$, blue line $b$ and $p$ axis only are
detected by  our corollary, but not by the theorem of Ref.\cite{QIC2010}. Similarly, these 6-nonseparable states in the
area enclosed   by purple line $c$, line $p=1$, green line $d$ and $p$ axis only are
detected by  our corollary, but not by the theorem of Ref.\cite{QIC2010}.}
  \end{center}
\end{figure}

$\emph{Example 2.}$ Discuss the $16$ qubit mixed state
\begin{eqnarray*}
\rho(p)=p|\chi\rangle\langle \chi|+\frac{1-p}{2^{16}}\textbf{1},
\end{eqnarray*}
where $|\chi\rangle=\dfrac{1}{3\sqrt{715}}\sum\limits_{l_1+l_2+\cdots +l_{15}=7}|1\rangle|l_1l_2\cdots l_{15}\rangle$
with $l_1,l_2,\cdots,l_{15}=0,1$.

In our corollary, let $\{A_i^{(1)},\cdots,A_i^{(n)}\}=\{\sigma_x,\sigma_y,\sigma_z\} $ and $A_i=\sigma_z$ for $1\leq i\leq 16$.
For $1\leq k\leq10$, these quantum states $\rho(p)$ contain $(k+1)$-partite entanglement detected by our corollary when $p_k< p\leq1$,
by inequality (14) in Ref.\cite{Hyllus2012} when $p_k'< p\leq1$. The specific values for $p_k$ and $p_k'$ are given in the table I.
For $p_k<p\leq p'_k$, these quantum states containing $(k+1)$-partite entanglement only can be tested by our corollary, but not by inequality (14) in Ref.\cite{Hyllus2012}. It's worth noting that inequality (14) of Ref.\cite{Hyllus2012} cannot identify any 10-partite entanglement and 11-partite entanglement.

\begin{table}
\caption{\label{tab:table1}
For $\rho(p)=p|\chi\rangle\langle \chi|+\dfrac{1-p}{2^{16}}\textbf{1}$, the thresholds of $p_k$, $p'_k$  for  the quantum states containing $(k+1)$-partite entanglement detected by our corollary and
inequality (14) in Ref.\cite{Hyllus2012} are illustrated, respectively.
When $p_k<p\leq1$ and $p'_k<p\leq1$, $\rho(p)$ contains $(k+1)$-partite entanglement detected by our corollary and
inequality (14) in Ref.\cite{Hyllus2012}, respectively.   For $p_k< p\leq p'_k$, these quantum states contain $(k+1)$-partite entanglement detected only by our corollary, but not by inequality (14) in Ref.\cite{Hyllus2012}. The symbol $\setminus$ means that inequality (14) of Ref.\cite{Hyllus2012} cannot identify any 10-partite entanglement and 11-partite entanglement.
}
\begin{ruledtabular}
\begin{tabular}{ccccccccccc}
$k$&1&2&3&4&5&6&7&8&9&10\\
\hline
&&&&&&&&&&\\
 $p_k$ & 0.1245 & 0.1401 & 0.1712 & 0.2179 & 0.2802 & 0.3580 & 0.4513 & 0.5603 & 0.7003 & 0.8560 \\
&&&&&&&&&&\\
   $p_k'$ & 0.1245 & 0.2491 & 0.3580 & 0.4981 & 0.5915 & 0.6848 & 0.7938 & 0.9961 & $\setminus$ & $\setminus$
\end{tabular}
\end{ruledtabular}
\end{table}

\section{Conclusions}
In this paper, we have investigated the characterization of $k$-nonseparability and $k$-partite entanglement from the perspective of quantum Fisher information. Effective criteria to identify $k$-nonseparability and $k$-partite entanglement of arbitrary-dimensional multipartite quantum systems are presented. Moreover, we have illustrated the power of these criteria through several specific examples.

\begin{center}
{\bf ACKNOWLEDGMENTS}
\end{center}

This work was supported by  the National Natural Science Foundation of China under Grant Nos. 12071110, 11701135 and 11947073, the Hebei Natural Science Foundation of China under Grant No. A2020205014 and No. A2017403025, the Education Department of Hebei Province Natural Science Foundation under Grant No.  ZD2020167, and the Foundation of  Hebei GEO University under Grant No. BQ201615.

\end{document}